\documentclass{nature}

\usepackage{graphicx,epsfig}
\usepackage{amsmath,amssymb,mathtools}
\usepackage[rgb]{xcolor}
\usepackage{hyperref}

\makeatletter
\let\saved@includegraphics\includegraphics
\AtBeginDocument{\let\includegraphics\saved@includegraphics}
\renewenvironment*{figure}{\@float{figure}}{\end@float}
\renewenvironment*{table}{\@float{table}}{\end@float}
\makeatother

\title{Simplicial models of social contagion}

\author{Iacopo Iacopini$^{1,2}$, Giovanni Petri$^{3,4}$, Alain Barrat$^{5,3}$ \& Vito Latora$^{1,2,6,7}$}

\begin{document}

\maketitle

\begin{affiliations}
 \item School of Mathematical Sciences, Queen Mary University of London, London E1 4NS, United Kingdom
 \item The Alan Turing Institute, The British Library, London NW1 2DB, United Kingdom
 \item ISI Foundation, Via Chisola 5, 10126 Turin, Italy
 \item ISI Global Science Foundation,33 W 42nd St10036 New York NY, United States
 \item Aix Marseille Univ, Universit\'e de Toulon, CNRS, CPT, Marseille, France
 \item Dipartimento di Fisica ed Astronomia, Universit\`a di Catania and INFN, I-95123 Catania, Italy
 \item Complexity Science Hub Vienna (CSHV), Vienna, Austria
\end{affiliations}

\newpage

\section*{Abstract}

\begin{abstract}
Complex networks have been successfully used to describe the spread of diseases in populations of interacting individuals. Conversely, pairwise interactions are often not enough to characterize social contagion processes such as opinion formation or the adoption of novelties, where complex mechanisms of influence and reinforcement are at work. Here we introduce a higher-order model of social contagion in which a social system is represented by a simplicial complex and contagion can occur through interactions in groups of different sizes. Numerical simulations of the model on both empirical and synthetic simplicial complexes highlight the emergence of novel phenomena such as a discontinuous transition induced by higher-order interactions. We show analytically that the transition is discontinuous and that a bistable region appears where healthy and endemic states co-exist. Our results help explain why critical masses are required to initiate social changes and contribute to the understanding of higher-order interactions in complex systems.

 \end{abstract}

\newpage

\section*{Introduction}

Complex networks describe well the connectivity of systems of various
nature \cite{albert2002statistical,latora_nicosia_russo_2017} and are
widely used as the underlying -- and possibly multilayered \cite{radicchi2013abrupt}-- social structure on
which dynamical processes \cite{porter2006dynamical,barrat2008dynamical}, such as disease spreading
\cite{pastor2015epidemic}, diffusion and adoption of innovation \cite{valente1996network, cowan2004network, iacopini2018network}, and opinion formation \cite{watts2007influentials} occur.
For example, when modelling an epidemic spreading in a population \cite{pastor2015epidemic}, the transmission between infectious and healthy individuals is typically assumed: \emph{(i)} to occur through pairwise interactions between infectious and healthy individuals, and \emph{(ii)} to be caused even by a single exposure of a healthy individual to an infectious one. Such processes of {simple contagion} 
can be conveniently represented by transmission mechanisms along  the links of the network of contacts between individuals.

When dealing instead with social contagion phenomena,  such as the adoption of norms, behaviours or new products, or the diffusion of rumors or fads, the situation is more complex.  Simple epidemic-like contagion can suffice to describe some cases, such as easily convincing rumors or domino effects \cite{centola2007complex}.
In other situations, however, they do not provide a satisfactory description, especially in those cases where more complex dynamics of peer influence and reinforcement mechanisms are at work \cite{guilbeault2018complex}.
{Complex contagion} mechanisms have been proposed to account for these effects. 
As defined by Centola \& Macy \cite{centola2007complex}: ``a contagion is complex if its transmission requires an individual to have contact with two or more sources of activation'', i.e. if a ``contact with a single active neighbor is not enough to trigger adoption''. Complex contagion can hence be broadly defined as a process in which exposure to multiple sources presenting the same stimulus is needed for the contagion to occur.
Empirical evidence that contagion processes including multiple exposure can be needed to  describe social contagion has been provided in various contexts and experiments \cite{centola2010spread,ugander2012structural,weng2012competition,karsai2014complex,monsted2017evidence}.

Modelling of social contagion processes has been driven by these considerations in several directions. 
Threshold models assume that, in order to adopt a
novel behavior, an individual needs to be convinced by a fraction of his/her
social contacts larger than a given threshold
\cite{watts2002simple,centola2007complex,melnik2013multi,karsai2014complex,ruan2015kinetics,czaplicka2016competition}. 
The processes considered in such models 
are usually deterministic.
Another modelling framework for social contagion 
relies instead on generalizations of epidemic-like 
processes, with stochastic contagion processes 
whose rates might depend on the number of sources
of exposure to which an individual is linked to, i.e., with 
a complex contagion flavor
\cite{weng2012competition,cozzo2013contact,hodas2014simple,herrera2015understanding,o2015mathematical,czaplicka2016competition,tuzon2018continuous}.
All these models are however still defined on networks of interactions between individuals: even when multiple interactions are needed for a contagion to take place, in both threshold and epidemic-like models, the fundamental building blocks of the system are pairwise interactions, structurally represented by the links of the network on which the process is taking place.

Here, we propose to go further and take into account that 
contagion can occur in different ways, either through pairwise
interactions (the links of a network) or through group interactions, i.e., through 
higher-order structures. Indeed, while an individual can be convinced independently by each of his/her neighbors (simple contagion), or by the successive
exposure to the arguments of different neighbors (complex contagion),
a fundamentally different mechanism is at work if the neighbors of an
individual convince him/her {in a group interaction.}  For example, we can adopt a new norm because of two-body processes, 
which means we can get convinced, separately, by
each one of our first neighbors in our social network who have already adopted the norm.  However, this is qualitatively different from a mechanism of contagion in which we get convinced because we are part of a social group of three
individuals, and our two neighbors are both adopters. In this case the
contagion is a three-body process, which mimics the simplest multiple
source of reinforcement that induces adoption. The same argument can
easily be generalized to larger group sizes.

\begin{figure}
	\centering
	\includegraphics[width=0.99\textwidth]{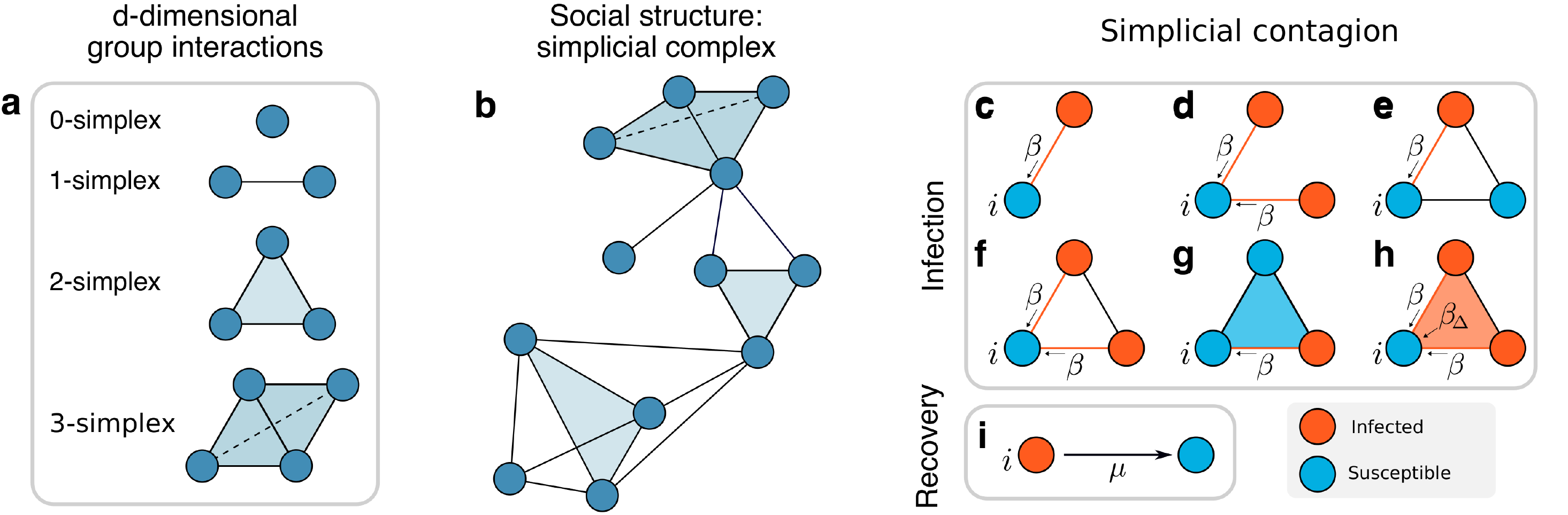}
	\caption{\label{fig:simplagion_model_viz}  {\bf Simplicial Contagion Model (SCM)}. The underlying structure of a social 
	system is made of simplices, representing 
	d-dimensional group interactions ({\bfseries\sffamily a}), organized in a simplicial complex ({\bfseries\sffamily b}). {\bfseries\sffamily c-h} Different channels of infection for a susceptible node $i$ in the Simplicial Contagion Model (SCM) of order $D=2$. Susceptible and infected nodes are colored in blue and red, respectively. Node $i$ is in contact with one ({\bfseries\sffamily c}, {\bfseries\sffamily e}) or more 
	({\bfseries\sffamily d}, {\bfseries\sffamily f})
	infected nodes  through links ($1$-simplices), and it becomes 
	infected with probability $\beta$ at each timestep through each of these links. {\bfseries\sffamily g-h} Node $i$ belongs to a 2-simplex (triangle). In {\bfseries\sffamily g} one of the nodes of the 2-simplex is not infected, so $i$ can only receive the infection from the (red) link, with probability $\beta$. In {\bfseries\sffamily h} the two other nodes of the $2$-simplex are infected, so $i$ can get the infection from each of the two $1$-faces (links) of the simplex with probability $\beta$, and also from the $2$-face with probability $\beta_2=\beta_{\Delta}$. {\bfseries\sffamily i} Infected nodes recover with probability $\mu$ at each timestep, as in the standard SIS model.}
\end{figure}

To build a modelling framework  based on these ideas, we formalise a social group as a simplex, and we adopt simplicial complexes as the underlying
structure of the social system under consideration (see Figure~\ref{fig:simplagion_model_viz}a-b). This simplicial representations is indeed more suited than networks to describe the co-existence of pairwise and higher-order interactions. We recall that, in its most basic definition, a $k$-simplex $\sigma$ is a set of $k+1$ vertices $\sigma = [p_0, \ldots, p_{k}]$. It is then 
easy to see the difference between a group interaction among three elements, which can be represented as a 2-simplex or ``full'' triangle $[p_0, p_1, p_2]$, and the collection
of its edges, $[p_0,p_1],[p_0,p_2],[p_1,p_2]$. Just like a collection 
of edges defines a network, a collection of simplices defines a simplicial complex. Formally, a simplicial
complex $\cal K$ on a given set of vertices $\mathcal{V}$, with
$|\mathcal{V}|=N$, is a collection of simplices, with the 
extra requirement that if 
simplex $\sigma \in \cal K$, then all the sub-simplices $\nu \subset
\sigma$ built from subsets of $\sigma$ are also contained in $\cal K$. 
Such a requirement, which makes 
simplicial complexes a special type of hypergraphs (see Supplementary Note 4), 
seems appropriate in the definition of higher-dimensional groups in the context 
of social systems, and simplicial complexes have indeed been used to represent social aggregation in human communication \cite{kee2013social}.
Removing this extra requirement would imply, for instance, modelling a group interaction of three individuals without taking into account also the 
dyadic interactions among them. The same argument can be extended to interactions of four or more individuals: it is reasonable to assume that  the existence of high-order interactions implies the presence of the lower-order interactions. 
For simplicity and coherence with the standard
network nomenclature, we call nodes (or vertices) the
$0$-simplices and links (or edges) the $1$-simplices
of a simplicial complex $\cal K$, while $2$-simplices correspond to the (``full") triangles, $3$-simplices to
the tetrahedra of $\cal K$, and so on (see Figure~\ref{fig:simplagion_model_viz}a).        
Simplicial complexes, differently from networks, can thus efficiently
characterize interactions between any number of units \cite{hatcher2002algebraic,salnikov2018simplicial}. Simplicial
complexes are not a new idea \cite{aleksandrov1998combinatorial}, but
the interest in them has been renewed
\cite{salnikov2018simplicial,sizemore2018importance,lambiotte2019networks}
thanks to the availability of new data sets and of recent advances in
topological data analysis techniques \cite{carlsson2009topology}. In
particular, they recently proved to be useful in
describing the architecture of complex networks
\cite{petri2013topological, sizemore2016classification, giles2019beyond} functional
\cite{petri2014homological, lord2016insights, lee2012persistent} and
structural brain networks \cite{sizemore2018cliques}, protein
interactions \cite{estrada2018centralities}, semantic networks
\cite{sizemore2018knowledge}, and co-authorship networks in science~\cite{patania2017shape}.

Here, we thus propose a new modelling framework for social contagion,
namely a model of {``simplicial contagion''}: this epidemic-like model
of social contagion on simplicial complexes takes into account the
fact that contagion processes occurring through a link or through a
group interaction both exist and have different rates.  Our model
therefore combines stochastic processes of simple contagion (pairwise
interactions) and of complex contagion occurring through group
interactions in which an individual is simultaneously exposed to
multiple sources of contagion.  We perform extensive numerical
simulations on both empirical data and synthetic simplicial complexes
and develop as well an analytical approach in which we derive and
solve the mean-field equations describing the evolution of density of
infected nodes.  We show both numerically and analytically that the
higher-order interactions lead to the emergence of new phenomena,
changing the nature of the transition at the epidemic threshold from
continuous to discontinuous and leading to the appearance of a
bistable region of the parameter space where both healthy and endemic
asymptotic states co-exist. The mean-field analytical approach
correctly predicts the steady-state dynamics, the position and the
nature of the transition and the location of the bistable region.  We
also show that, in the bistable region, a critical mass is needed to
reach the endemic state, reminding of the recently observed minimal
size of committed minorities required to initiate social
changes\cite{centola2018experimental}.

\section*{Results}

\subsection{The contagion model.}

In order to model a simplicial contagion process, we associate a
dynamical binary state variable $x$ to each of the $N$ vertices of $\cal K$, such that $x_i (t)\in\{0,1\}$ represents the state of vertex $i$ at time $t$. 
Using a standard notation, we divide the population of individuals into two classes of susceptible (S) and infectious (I) nodes, 
corresponding respectively to the values $0$ and $1$ of the state variable $x$.  In the context of adoption 
processes, the state I represents individuals who have adopted a behaviour.
At each time $t$, the macroscopic order parameter is given by the density of infectious nodes $\rho(t) = \frac{1}{N}\sum_{i=1}^{N}x_i(t)$.
The model we propose here, the so-called {Simplicial Contagion
Model} (SCM) {of order $D$}, with $D \in [1,N-1]$, is governed
by a set of $D$ control parameters $B = \{\beta_1, \beta_2, \dots,
\beta_{D}\}$, whose elements represent the probability per unit time
for a susceptible node $i$ that participates to a simplex $\sigma$ of
dimension $D$ to get the infection from each one of the subfaces
composing $\sigma$, under the condition
that all the other nodes of the subface are infectious. 
In practice, with this notation, $\beta_1$ is
equal to the standard probability of infection $\beta$ that a
susceptible node $i$ gets the infection from an infected neighbor $j$
through the link $(i,j)$ (corresponding to the process $S+I \to 2I$).
Similarly, the second parameter $\beta_2 \equiv \beta_{\Delta}$
corresponds to the probability per unit time that node $i$ receives
the infection from a ``full" triangle (2-simplex) $(i,j,k)$ in which
both $j$ and $k$ are infectious, $\beta_3=\beta_{\boxtimes}$ from
a group of size 4 (3-simplex) to which it belongs, and so on. Such
processes can be represented as $Simp(S,nI) \to Simp((n+1)I)$: a
susceptible node, part of a simplex of $n+1$ nodes among which all
other $n$ nodes are infectious, becomes infectious with probability per
unit time $\beta_n$. Thanks to the simplicial complex
requirements that all subsimplices of a simplex are included,
contagion processes in a $n$-simplex among which $p < n$ nodes are
infectious are also automatically considered, each of the $n+1-p$
susceptible nodes being in a simplex of size $p+1$ with the $p$
infectious ones. 
Notice, however, that this 
assumption can be dropped and the contagion model extended to the case of hypergraphs \cite{berge1984hypergraphs, ghoshal2009random} (see Supplementary Note 4).
Figure~\ref{fig:simplagion_model_viz}c-h illustrates
the concrete example of the six possible ways
in which a susceptible node $i$ can undergo social
contagion for an SCM of order $D=2$ with parameters $\beta$ and
$\beta_{\Delta}$.  Finally, the recovery dynamics ($I \to S$) is controlled by the node-independent recovery probability $\mu$ 
(Figure~\ref{fig:simplagion_model_viz}i). Notice that
the SCM of order $D$ reduces to the standard SIS model on a network when $D=1$, since in this case the infection can only be transmitted through the links of $\cal K$.

\subsection{Simplicial contagion on real-world simplicial complexes.} 

To explore the phenomenology of the simplicial contagion model, we first consider its evolution
on empirical social structures. To this aim, we 
consider publicly available data sets describing 
face-to-face interactions collected by the SocioPatterns collaboration \cite{sociopatterns}.
Face-to-face interactions represent indeed
a typical example in which group encounters 
are fundamentally different from sets of 
binary interactions and can naturally be 
encoded as simplices. The time-resolved
nature of the data allows us to create simplicial
complexes describing the aggregated social structure, as described in Methods. For
simplicity, we only consider simplices of
dimension up to $D=2$. We consider data on 
interactions collected in four different 
social contexts: a workplace, a conference, a hospital and a high school (see Methods for
details on the data sets).

We simulate the SCM over the 
simplicial complexes obtained from the four data sets as described in Methods. 
In particular, we start with an initial density $\rho_0$ of infectious nodes and we run the simulations by taking into consideration all the possible channels of infection illustrated in Figure~\ref{fig:simplagion_model_viz}c-h.
We stop a simulation if an absorbing state is reached, otherwise we compute the average stationary density of infectious nodes 
$\rho^*$ by averaging the values measured
in the last $100$ time-steps after reaching a stationary state. 
The results are averaged over $120$ runs obtained with  
randomly placed initial infectious nodes 
with the same density $\rho_0$.
Moreover, the different data sets correspond to different
densities of $1$- and $2$-simplices
(see Supplementary Note 1). We thus
rescale the infectivity parameters
$\beta$ and $\beta_\Delta$ respectively
by the average degree $\langle k \rangle$
and by the average number of $2$-simplices
incident on a node, $\langle k_\Delta \rangle$.
We finally express all
results as functions of the rescaled parameters 
$\lambda=\beta\langle k \rangle/\mu$  and $\lambda_{\Delta}= \beta_\Delta \langle k_\Delta \rangle/\mu$.

\begin{figure}
	\centering
	\includegraphics[width=0.6\textwidth]{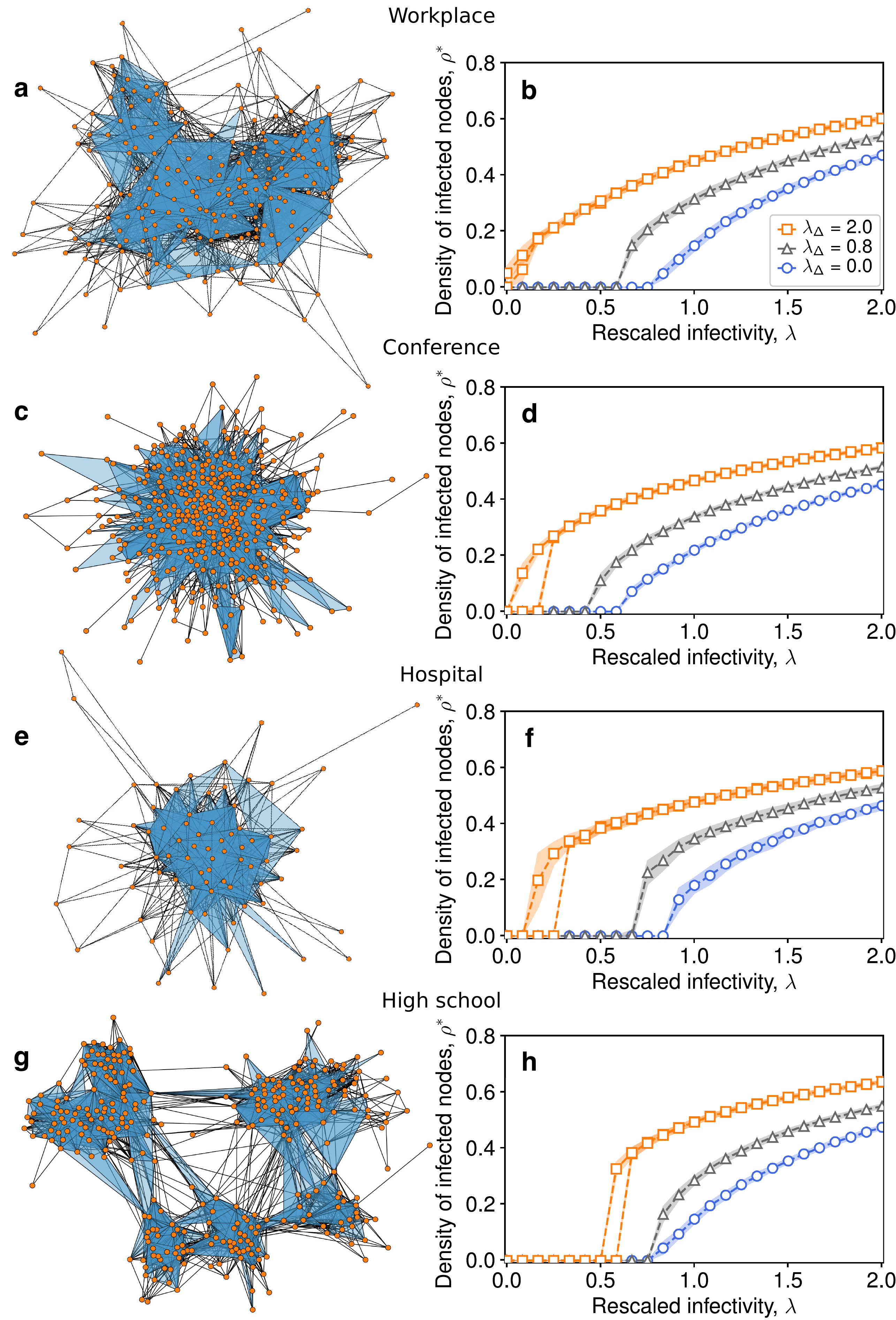}
	\vspace{-1em}
	\caption{\label{fig:simplagion_sociopatterns} {\bf SCM of order $D=2$ on real world higher-order social structures.} Simplicial complexes are constructed from high-resolution face-to-face contact data recorded in four different context: {\bfseries\sffamily a} a workplace, {\bfseries\sffamily c} a conference, {\bfseries\sffamily e} a hospital and {\bfseries\sffamily g} a high school. Prevalence curves are respectively reported in panels {\bfseries\sffamily b}, {\bfseries\sffamily d}, {\bfseries\sffamily f} and {\bfseries\sffamily h}, in which the average fraction of infectious nodes obtained in the numerical simulations is plotted against the rescaled infectivity $\lambda=\beta\langle k \rangle/\mu$ for different values of the 
	rescaled parameter $\lambda_{\Delta}= \beta_\Delta \langle k_\Delta \rangle/\mu$, namely 
	$\lambda_{\Delta}=0.8$ (black triangles) and $\lambda_{\Delta}=2$ (orange squares). The blue circles denote the simulated curve for the equivalent standard SIS model ($\lambda_{\Delta}=0$), which does not consider higher order effects. For $\lambda_{\Delta}=2$ a bi-stable region appears, where healthy and endemic states co-exist.}
\end{figure}

Figure~\ref{fig:simplagion_sociopatterns} shows the 
resulting prevalence curves for the four data sets
(see also Supplementary
Note 5). 
In each panel (b,d,f,h), the average fraction of
infected nodes $\rho^*$ in the 
stationary state is plotted as a function of the rescaled infectivity $\lambda=\beta\langle k \rangle/\mu$  for simulations of the SCM with $\lambda_{\Delta}=0.8$ (black triangles) and $\lambda_{\Delta}=2$ (orange squares).
For comparison, we also plot the case $\lambda_{\Delta}=0$,
which is equivalent to the standard SIS model with
no higher-order effects (blue circles).
We observe two radically different behaviours 
for the two values of $\lambda_\Delta \neq 0$. 
For $\lambda_{\Delta}=0.8$, the density of infectious nodes varies as a function of $\lambda$ in a very similar way to the case $\lambda_\Delta=0$
(simple contagion), with a continuous transition. 
For $\lambda_\Delta =2$ we observe instead 
the appearance of an endemic state with $\rho^* > 0$ at a value of $\lambda^c$ well below the epidemic threshold of the other two cases. 
Furthermore, this transition appears to be discontinuous, and an hysteresis loop appears 
in a bi-stable region, where both healthy $\rho^* = 0$ and endemic $\rho^* > 0$ states can co-exist (dashed orange lines): in this parameter
region, the final state
depends on the initial density of
infectious nodes $\rho_0$. 

The simplicial complexes
used in these simulations correspond to 
various social contexts and different densities
of 1- and 2-simplices, and yield a similar phenomenology. These empirical structures 
however exhibit distributions of generalized 
degrees that are not well peaked around their average values (see Supplementary Note 1), and
do not allow us to systematically explore
size effects. 
To better understand the phenomenology of the simplicial contagion model, we thus now 
explore its behaviour on 
synthetic simplicial complexes with controlled properties.

\subsection{Simplicial contagion on synthetic simplicial complexes.}

A range of models for random simplicial complexes have been proposed so far, starting from the exponential random simplicial complex, the growing and generalized canonical ensemble \cite{zuev2015exponential, courtney2016generalized, courtney2017weighted} and the simplicial configuration models \cite{young2017construction} to the simplicial activity driven model \cite{petri2018simplicial} 
generalizing the activity driven temporal network model~\cite{perra2012activity}. 
While these yield Erd\"os-R\'enyi-like models \cite{kahle2009topology, costa2016random} of arbitrary complexity, here we are interested in models generating simplicial complexes with simplices of different dimension in which we can control and tune the expected local connectivity, e.g. the number of edges and ``full" triangles a node belongs to.
We therefore propose a new model to construct 
random simplicial complexes, the RSC model, which allows us to maintain the average
degree of the nodes, $\langle k_1 \rangle$, 
fixed, while varying at the same time the expected number of ``full" triangles (2-simplices) $\langle k_\Delta \rangle$ incident on a node.
The RSC model of dimension $D$ has $D+1$ parameters, namely the number of vertices $N$ and $D$ probabilities $\{p_1,\dots,p_k,\dots,p_D\}$, $p_k \in [0,1]$, which control for the creation of $k$-simplices up to dimension $D$. For the purpose of this study we limit the RSC model to $D=2$, which restricts the set of required parameters to $(N,p_1,p_2)$, but the procedure could easily be extended to larger $D$. The model works as follows. We first create 1-simplices (links) as in the Erd\"os-R\'enyi model \cite{erdos1960evolution}, by connecting any pair $(i,j)$ of vertices with probability $p_1$. Similarly, 2-simplices are then created by connecting any triplet $(i,j,k)$ of vertices with probability $p_2\equiv p_{\Delta}$. Notice that simplicial complexes built in this way are radically different from the clique complexes obtained from Erd\"os-R\'enyi graphs\cite{kahle2009topology}, in which every subset of nodes forming a clique is automaticaly ``promoted" to a simplex. Contrarily, in a simplicial complex generated by the RSC model proposed here, a 2-simplex $(i,j,k)$ does not come from the promotion of an ``empty" triangle composed by three 1-simplices $(i,j)$, $(j,k)$, $(k,i)$ to a ``full triangle" $(i,j,k)$. This also means that the model allows for the presence of $(k+1)$-cliques that are not considered $k$-simplices, therefore it is able generate simplicial complexes having both ``empty" and ``full" triangles, respectively encoding three 2-body interactions and one 3-body interactions (as for instance in Fig.~\ref{fig:simplagion_model_viz}b).
The expected average numbers of 1- and 2-simplices incident on a node, noted
$\langle k \rangle$ and $\langle k_{\Delta} \rangle$,
are easy to calculate (see Methods). Therefore, for any given size $N$, we can produce simplicial complexes having desired values of $\langle k \rangle$ and $\langle k_{\Delta} \rangle$ by appropriately tuning $p_1$ and $p_\Delta$. More details about the construction of the model and the tuning of the parameters are provided in the ``Methods'' section, while the agreement between the expected values of $\langle k \rangle$ and $\langle k_{\Delta} \rangle$ with the empirical averages obtained from different realizations of the model is discussed in Supplementary Note 1.

\begin{figure}
	\centering
	\includegraphics[width=\textwidth]{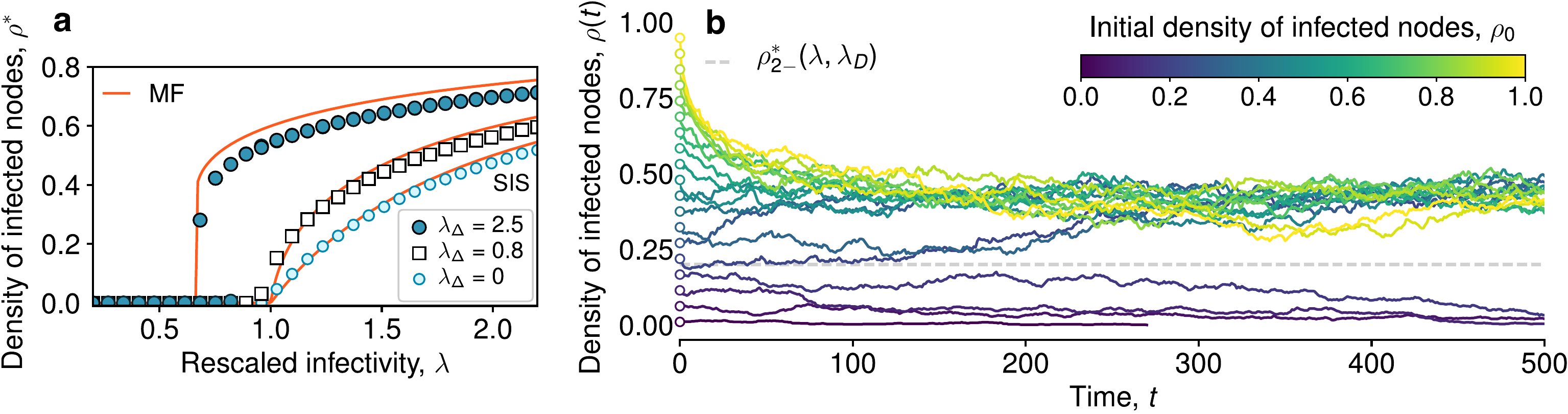}
	\caption{\label{fig:simplagion_RSC} {\bf SCM of order $D=2$ on a
		synthetic random simplicial complex (RSC).} The RSC is generated with
		the procedure described in this manuscript, with parameters
		$N=2000$, $p_1$ and $p_{\Delta}$ tuned in order to produce a
		simplicial complex with $\langle k \rangle \sim 20$ and $\langle
		k_{\Delta} \rangle \sim 6$. {\bfseries\sffamily a} The average fraction of infected obtained by means of numerical simulations is plotted against the
		rescaled infectivity $\lambda=\beta\langle k \rangle/\mu$ for
		$\lambda_{\Delta}=0.8$ (white squares) and $\lambda_{\Delta}=2.5$
		(filled blue circles). The light blue circles give the numerical results for the
		standard SIS model ($\lambda_{\Delta}=0$) that does not
		consider higher order effects. 
		The red lines correspond to the
		analytical mean field solution described by
		Equation~\eqref{eq:MF_SIS_d2_bis}. For $\lambda_{\Delta}=2.5$ we observe
		a discontinuous transition with the formation of a bistable
		region where healthy and endemic states co-exist.
		{\bfseries\sffamily b} Effect of the initial density of infected nodes, shown by the temporal evolution of the densities of infectious nodes (a single realization is shown for each value of the initial density). The infectivity parameters are set within the range in which we observe a bistable region ($\lambda=\beta\langle k \rangle/\mu=0.75$, $\lambda_{\Delta} = \beta_{\Delta}\langle k_{\Delta} \rangle/\mu=2.5$). Different curves - and different colors - correspond to different values for the initial density of infectious nodes $\rho_0\equiv\rho(0)$. The dashed horizontal line corresponds to the unstable branch $\rho^*_{2-}$ of the mean field solution given by Equation~\ref{eq:SIS_MF_sol}, which separates the two basins of attraction.}
\end{figure}

We simulate the SCM over a RSC created with the procedure described above,
with $N=2000$ nodes, $\langle k \rangle \simeq 20$ and $\langle
k_{\Delta} \rangle \simeq 6$. As for the real-world  simplicial complexes,
we start with a seed of $\rho_0$ infectious nodes placed at random and
we compute the average stationary density of infectious $\rho^*$ by averaging over different runs, each one 
using a different instance of the RSC model.
Results are shown in Fig.\ref{fig:simplagion_RSC}a, where the average fraction of
infected nodes, as obtained by the simulations, is plotted as a function of the rescaled
infectivity $\lambda=\beta\langle k \rangle$ 
for a ($D=2$) SCM with
$\lambda_{\Delta}=0.8$ (white squares), $\lambda_{\Delta}=2.5$
(filled blue circles) and $\lambda_{\Delta}=0$ (light blue circles).

Despite the very different properties of the underlying structure, the dynamics of the SCM
on the RSC is very similar to the one observed on
the real-world simplicial complexes.
For $\lambda_{\Delta}=0.8$ the model behaves similarly to a simple contagion model
($\lambda_\Delta=0$), with a continuous transition at $\lambda^c=1$, the well-know epidemic threshold of the standard SIS model on homogeneous networks.
When a higher value of $\lambda_{\Delta}$ is considered ($\lambda_{\Delta}=2.5$), the epidemic can be sustained below $\lambda^c=1$, and 
both an epidemic-free and an endemic states are present in
the region $\lambda^c<\lambda<1$, with appearance of
a hysteresis loop (see the filled blue circles
in Fig.~\ref{fig:simplagion_RSC}a). 
In this region, we obtain
$\rho(t \to \infty)=0$ for $\rho(t=0)=0.01$, 
while $\rho(t \to \infty) > 0$ for $\rho(t=0)=0.4$.
The size-dependence of the hysteresis loop
is shown in Supplementary Note 2 to be very small.
The dependency from the initial conditions is 
also further illustrated in
Fig.~\ref{fig:simplagion_RSC}b, in which the temporal dynamics of single runs are shown. 
The various curves show how the density of infected nodes $\rho(t)$ evolves when initial seeds of 
infected nodes of different sizes are considered. Each color corresponds to a different value of $\rho_0$, with brighter colors representing higher initial densities of infected individuals. 
The figure clearly shows 
the presence of a threshold value for $\rho_0$, such
that $\rho(t)$ goes to the absorbing state $\rho(t)=0$ if $\rho_0$ is smaller
than the threshold, and to a non-trivial steady
state if the initial density is above the threshold.

\subsection{Mean field approach.}

In order to study more extensively this phenomenology as
$\lambda_\Delta$ and $\lambda$ vary, and to further characterize the discontinuous transition,
we consider a mean field (MF) description of the SCM, 
under a homogeneous mixing hypothesis \cite{kiss2017mathematics}.
Given the set of infection probabilities 
$B\equiv\{\beta_{\omega}, \omega=1, \cdots , D\}$ and 
a recovery probability $\mu$, 
we assume the independence
between the states $x_{i}(t)$ and $x_{j}(t)$ $\forall\ i, j \in \mathcal{V}$, and we
write a MF expression for the temporal evolution of the density of
infected nodes $\rho(t)$ as: 
\begin{equation}\label{eq:MF_SIS}
d_{t}\rho(t) = -\mu\rho(t) + \sum_{\omega=1}^{D} \beta_{\omega}\langle k_{\omega}\rangle \rho^{\omega}(t)\bigl[1-\rho(t)\bigr]  
\end{equation}
where, for each $\omega=1,\cdots,D$, 
$k_{\omega}(i)=k_{\omega,0}(i)$ is the generalized (simplicial) degree of a
$0$-dimensional face (node $i$), i. e., the number of
$\omega$-dimensional simplices incident to the node $i$
\cite{courtney2016generalized, courtney2017weighted}, and $\langle
k_{\omega}\rangle$ is its average over all the nodes $i\in
\mathcal{V}$. With this approximation we assume that the local
connectivity of the nodes is well described by globally averaged
properties, such as the average generalized degree.
We can immediately check that in the case $D=1$ we recover
the standard MF equation for the SIS model, which leads to the well
known stationary state solutions $\rho^{* \left[D=1\right]}_1=0$ and
$\rho^{* \left[D=1\right]}_2 = 1 - \mu/(\beta\langle k \rangle)$. The absorbing state
$\rho^{* \left[D=1\right]}_1=0$ is the only solution
for $\beta\langle k \rangle/\mu < 1$, 
i.e., below the epidemic threshold.  When 
$\beta\langle k \rangle /\mu > 1$, this state 
becomes unstable while the solution 
$\rho^{* \left[D=1\right]}_2$ becomes 
a stable fixed point
of the dynamics. The transition between these two
regimes is continuous at $\beta\langle k \rangle/\mu = 1$.

Let us now focus on a more interesting but still analytically tractable
case in which we extend the contagion dynamics up to dimension $D=2$,
so that Equation \eqref{eq:MF_SIS} reads:
\begin{equation}\label{eq:MF_SIS_d2}
d_{t}\rho(t) = -\mu\rho(t) + \beta \langle k\rangle \rho(t)\bigl[1-\rho(t)\bigr] 
+  \beta_{\Delta} \langle k_{\Delta}\rangle \rho^{2}(t)\bigl[1-\rho(t)\bigr]
\end{equation}
where $\langle k_{\Delta}\rangle \equiv \langle k_{2}\rangle$. 
By defining as before $\lambda= \beta \langle k\rangle/\mu$ and 
$\lambda_{\Delta}= \beta_{\Delta} \langle k_{\Delta}\rangle/\mu$, and by rescaling the time by $\mu$,
we can rewrite Equation \eqref{eq:MF_SIS_d2} as: 
\begin{equation}\label{eq:MF_SIS_d2_bis}
d_{t}\rho(t) = - \rho(t) 
(\rho(t) - \rho^{*}_{2+}) (\rho(t) - \rho^{*}_{2-}) \ ,
\end{equation}
where $\rho^{*}_{2+}$ and $\rho^{*}_{2-}$
are the solutions of the second order 
equation 
$1 -\lambda (1-\rho) - \lambda_\Delta \rho (1-\rho) = 0$.
We thus obtain: 
\begin{equation}\label{eq:SIS_MF_sol}
\rho^{*}_{2\pm} = \frac{ \lambda_{\Delta}-\lambda \pm \sqrt{(\lambda - \lambda_{\Delta})^2 -4\lambda_{\Delta}(1-\lambda)} }{2\lambda_{\Delta}} .
\end{equation}

The steady state equation $d_{t}\rho(t)=0$ has thus up to three solutions in the acceptable range $\rho \in [0,1]$. The 
solution  $\rho^{*}_1=0$  corresponds to the usual
absorbing epidemic-free state, in which all the individuals recover and the spreading dies out. 
A careful analysis of the stability of this state 
and of the two other solutions 
$\rho^{*}_{2+}$ and $\rho^{*}_{2-}$ is however
needed to fully characterize the phase diagram of the system.

Let us first consider the case $\lambda_\Delta \le 1$.
It is possible to show that $\rho^*_{2-}$, when it
is real-valued, is always
negative, i.e., it is not an acceptable solution. Moreover,
$\rho^*_{2+}$ is positive for $\lambda >1$ and negative
for $\lambda < 1$. In the regime $\lambda_\Delta \le 1$ therefore, if $\lambda < 1$, the only acceptable solution to $d_{t}\rho(t) = 0$ is  $\rho^{*}_1=0$; contrarily, for $\lambda > 1$, since $\rho^*_{2-} < 0$ and $\rho^*_{2+} > 0$,  Equation \eqref{eq:MF_SIS_d2_bis} shows that $d_{t}\rho(t)$ is positive at small $\rho(t)$: the absorbing state $\rho^{*}_1=0$ is thus unstable and the solution $\rho^*_{2+}$ is stable. As $\rho^*_{2+}=0$ for $\lambda=1$, the transition at the epidemic threshold $\lambda = 1$ is continuous.
In conclusion, when $\lambda_\Delta \le 1$, 
the transition is similar to the one of the standard SIS model
with $\lambda_\Delta = 0$.

Let us now consider the case of 
$\lambda_\Delta > 1$.
Then, for $\lambda < \lambda^c = 2 \sqrt{\lambda_\Delta} - \lambda_\Delta$, both $\rho^*_{2+}$ and $\rho^*_{2-}$
are outside the real domain, and the only
steady state is the absorbing one $\rho^{*}_1=0$.
Note that $\lambda^c < 1$, since $\lambda_\Delta > 1$.
For $\lambda > \lambda^c$, we thus have two possibilities
to consider. If $\lambda > 1$, we can show that $\rho^*_{2-} < 0 < \rho^*_{2+}$. Equation \eqref{eq:MF_SIS_d2_bis} shows then that, for small $\rho(t)$, $d_t \rho(t) > 0$: as above, the absorbing state $\rho^{*}_1=0$ is unstable and the density of infectious nodes tends to $\rho^*_{2+}$ in the large time limit; if instead $\lambda^c < \lambda < 1$, we obtain that $0 < \rho^*_{2-} < \rho_{2+}^*$. Then, still from Equation \eqref{eq:MF_SIS_d2_bis}, we obtain that $d_t \rho(t) < 0$ for $\rho(t)$ between $0$ and $\rho^*_{2-}$, and that $d_t \rho(t) > 0$ for $\rho(t)$ between $\rho^*_{2-}$ and $\rho^*_{2+}$. As a result, both  $\rho^{*}_1=0$ and $\rho_{2+}^*$ are stable steady states of the dynamics, while $\rho^*_{2-}$ is an unstable solution. Most interestingly, the long time limit of the dynamics depends then on the initial conditions. Indeed, if the initial density of infectious nodes, $\rho(t=0)$, is below $\rho^*_{2-}$, the short time derivative of $\rho(t)$ is negative, so that the density of infectious nodes decreases and the system tends to the absorbing state: $\rho(t) \xrightarrow[t \to \infty]{} 0$. On the other hand, if the initial density $\rho(t=0)$ is large enough (namely, larger than $\rho^*_{2-}$), the dynamical evolution Equation \eqref{eq:MF_SIS_d2_bis} pushes the density towards the value $\rho_{2+}^*$, i.e. $\rho(t) \xrightarrow[t \to \infty]{} \rho^*_{2+}$. Since $\rho_{2+}^* > 0$, the transition at $\lambda_c$ is discontinuous.

\begin{figure}
	\centering
	\includegraphics[width=0.99\textwidth]{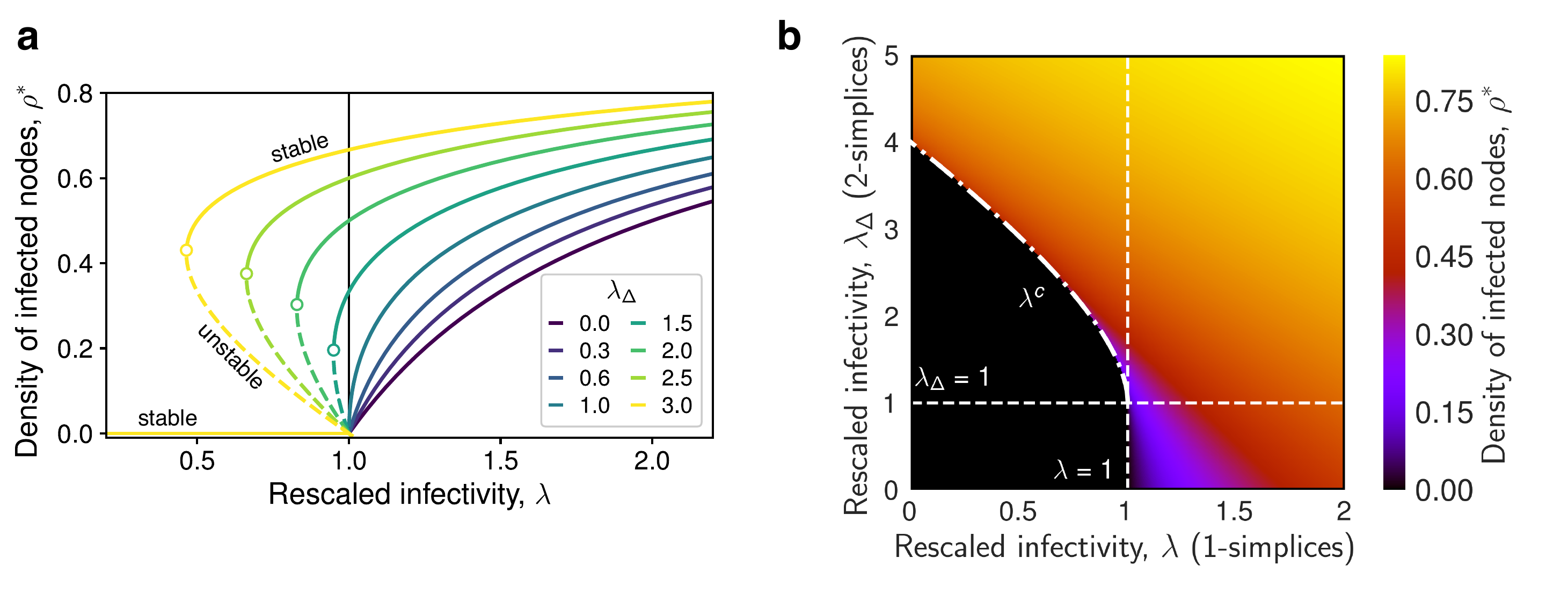}
	\vspace{-2em}
	\caption{\label{fig:MF} {\bf Phase diagram of the SCM of order $D=2$ in
		mean field approximation.} {\bfseries\sffamily a} The stationary solutions $\rho^{*}$ given by Equation~(\ref{eq:SIS_MF_sol}) are plotted as a function of the rescaled link infectivity $\lambda=\beta\langle k \rangle/\mu$. Different curves correspond to different values of the triangle infectivity
		$\lambda_{\Delta}=\beta_{\Delta}\langle
		k_{\Delta} \rangle/\mu$. Continuous and dashed lines correspond to stable and unstable branches respectively, while the vertical line denotes 
		the epidemic threshold $\lambda^c=1$ in the standard SIS
		model that does not consider higher order effects. For $\lambda_{\Delta}\le1$ the higher order interactions only contribute to an increase in the density of infected individuals in the endemic state, while they leave the threshold unchanged. Conversely, when $\lambda_{\Delta}>1$ we observe a shift of the epidemic threshold, and a change in the type of transition, which becomes discontinuous.
		{\bfseries\sffamily b} Heatmap of the stationary solution $\rho^{*}$ given by Equation~(\ref{eq:SIS_MF_sol}) as a
		function of the rescaled infectivities $\lambda=\beta\langle k
		\rangle/\mu$ and $\lambda_{\Delta}=\beta_{\Delta}\langle k_{\Delta}
		\rangle/\mu$. The black area corresponds
		to the values of $(\lambda, \lambda_\Delta)$ such that
		the only stable solution is $\rho_1^*=0$.
		The dashed vertical line corresponds to $\lambda=1$, the epidemic threshold of the standard SIS model without higher order effects. The dash-dotted line represents the points 
		$(\lambda^c, \lambda_\Delta)$, with $\lambda^c = 2\sqrt{\lambda_\Delta} - \lambda_\Delta$, where the system undergoes a discontinuous transition.}
\end{figure}

We illustrate these results by showing in Fig.~\ref{fig:MF}a the solutions $\rho_1^*$, $\rho_{2+}^*$ and $\rho^*_{2-}$
as a function of $\lambda$ and for different values of $\lambda_{\Delta}$. 
The vertical line corresponds to the standard epidemic threshold for the SIS model ($\lambda_\Delta=0$). Dashed lines depict unstable branches, as given by 
$\rho^{*}_{2-}$.
We emphasize again two important points. First,  for $\lambda_\Delta > 1$ we observe a discontinuous
transition at $\lambda^c = 2 \sqrt{\lambda_\Delta} - \lambda_\Delta$,
instead of the usual continuous transition at the epidemic threshold. Second,
for $\lambda^c < \lambda < 1$
the final state depends on the initial density of infectious {nodes}, 
as described above:
the absorbing state $\rho^*_1=0$ is reached if the 
initial density $\rho(t=0)$ is below the 
unstable steady state value $\rho^*_{2-}$; on the
contrary, if $\rho(t=0)$ is above this value, the system
tends to a finite density of infectious nodes
equal to $\rho_{2+}^*$.
In other words, a critical mass is needed
to reach the endemic state, reminding of the recently observed minimal
size of committed minorities required to initiate social changes 
\cite{centola2018experimental}.

Figure~\ref{fig:MF}b is a two-dimensional phase diagram showing $\rho^{*}_{2+}$ for different values of $\lambda$ and $\lambda_\Delta$. Lighter colours correspond to higher values of the stationary density of infectious nodes, while the dashed vertical line corresponds to the epidemic threshold of the standard (without higher order effects) SIS model, namely $\lambda=1$.
For  $\lambda_\Delta \le1$ (below the dashed horizontal line) the transition as $\lambda$ crosses $1$ is seen to be continuous, while, for $\lambda_\Delta >1$, the transition is clearly discontinuous along the curve $\lambda^c = 2\sqrt{\lambda_\Delta} - \lambda_\Delta$ (dash-dotted line).
The analytical values of $\rho_{2+}^*$ are also reported 
as continuous red lines in Figure~\ref{fig:simplagion_RSC}a and compared to the results of the simulations, showing in this way the accuracy of the mean field approach just described.
In addition, Figure~\ref{fig:simplagion_RSC}b shows that the unstable solution $\rho_{2-}^*$ accurately separates the two basins of attractions for the dynamics, i.e., it defines the critical initial density of infected $\rho_0$ that determines whether the long term dynamics reaches the healthy state or the endemic one.
Notice that the mean field approach is in fact able to correctly capture both the position of the thresholds and the discontinuous 
nature of the transition for the SCM with $\lambda_\Delta > 1$.

We finally note that, while a general solution for
general $D$ with arbitrary parameters $\{\beta_\omega\}$ 
remains out of reach, it is possible to show
that the phenomenology obtained
for $D=2$ is also observed for specific cases
with $D \ge 3$. We consider indeed in the Supplementary
Note 3 two cases: $D=3$ with $\beta_2=0$ and
general $D > 3$ with $\beta_1 = \cdots = \beta_{D-1}=0$. 
In both cases, we show the appearance of a discontinuous
transition in the regime where the simple contagion
$\beta_1$ is below threshold (i.e., $\beta_1 \langle k \rangle < \mu$): 
similarly to the case $D=2$,
this transition occurs
as $\beta_D$, which describes the rate
of the high-order contagion process, increases.

\subsection{Discussion.}

In summary, the simplicial model of contagion introduced in
this work is able to capture the basic mechanisms and effects of
higher-order interactions in social contagion processes.  Our analytical results were derived in a mean field approximation
and indeed quantatively compared to 
the nondescript simplicial complexes obtained
in our random simplicial complex model
(akin to ER simplicial complexes\cite{costa2016random}). However,
the framework we introduced is very general
and the phenomenology robust, as seen from the
results obtained on empirical data sets.
It would be interesting to investigate the SCM on
more general 
simplicial complexes with for instance heterogeneous generalized degree distribution or with community structures, and to consider simplicial complexes with emergent properties such as hyperbolic geometry \cite{bianconi2017emergent,mulder2018network,bianconi2018topological}, or
temporally evolving simplicial complexes \cite{petri2018simplicial}.
Furthermore, given that the SCM can be mapped on a model with hypergraphs if the hyperedges of different types are carefully chosen, it would be interesting to study the behavior of complex contagion processes on more general classes of hypergraphs \cite{bodo2016sis,lanchier2013stochastic}.
Finally, we hope that the idea will be extended from spreading processes
to other dynamical systems, for instance to Kuramoto-like models with
higher-order terms. Developing and studying such systems might allow to better take into account higher-order dynamical effects in real data-driven models.

\begin{methods}

\subsection{Data description and processing.}

We consider four data sets of face-to-face interactions collected in different social contexts: a workplace (InVS15)~\cite{genois2018can}, a conference (SFHH)~\cite{isella2011s}, a hospital (LH10)~\cite{vanhems2013estimating} and a high school (Thiers13)~\cite{mastrandrea2015contact}. In each case face-to-face interactions have been measured with a temporal resolution of 20 seconds. We first aggregated the data by using a temporal window of $\Delta t=5$ minutes, and computed all the maximal cliques that appear. Since we limit our study to the case $D=2$, we need to produce a clique complex formed by 1- and 2-simplices. Therefore, we considered all the 2- and 3-cliques and weight them according to their frequency. Note that while higher-dimensional cliques are not included in the final simplicial complex, their sub-cliques up to size 3 are considered in the counting. We then retained $20\%$ of the simplices with the largest number of appearances.
The thresholded simplicial complexes obtained in this way are those used in Supplementary Figure 6. Their connectivity properties are summarised in Table \ref{table:sociopatterns_data}.

To reduce finite size effects, we augmented the thresholded simplicial complexes as follows: for each data set we extracted the list of sizes of the maximal simplices, also called facets, and the list of pure simplicial degrees of nodes. 
We then duplicated these lists five times and used the extended lists as input for the simplicial configuration model, described in Ref.~\cite{young2017construction}. 
The outputs of this procedure are simplicial complexes with the same statistical properties as the input complex but of significantly larger size. We used these augmented complexes as substrates for the simulations shown in Figure \ref{fig:simplagion_sociopatterns}.

\begin{table} 
	\begin{center} 
		\begin{tabular}{|l l|c c|c c|} 
			\hline
			Dataset & Context & $\langle k\rangle$ & $\langle k_\Delta \rangle$ & $\langle k\rangle^{\text{aug}}$ & $\langle k_\Delta \rangle^{\text{aug}}$\\
			\hline
			InVS15 & Workplace &16.9 &7.0 &21.0 &7.0  \\
			SFHH & Conference &15.0 &7.6 &21.6 &7.7  \\
			LH10 & Hospital &19.1 &17.1 &25.7 &17.5  \\
			Thiers13 & High school &20.1 &10.9 &32.0 &11.1  \\
			\hline
		\end{tabular}
		\caption[]{{\bf Statistics of real-world simplicial complexes.} Average generalized degree of the four real-world  simplicial complexes constructed from the considered data sets (before and after the data augmentation).}
		\label{table:sociopatterns_data}
	\end{center}
\end{table}

\subsection{Construction of random simplicial complexes.}

The random simplicial complex (RSC) model produces simplicial complexes of dimension $D=2$  as follows. Given a set $\mathcal{V}$ of $N$ vertices we connect any two nodes $i,j\in \mathcal{V}$ with probability $p_1 \in [0,1]$, so that the average degree, at this stage, is $(N-1)p_1$. Then, for any $i, j, k \in\mathcal{V}$, we add a 2-simplex $(i,j,k)$ with 
probability $p_{\Delta}\in [0,1]$.
At this point each node has an average number $\langle k_{\Delta} \rangle = (N-1)(N-2)p_{\Delta}/2$ of incident 2-simplices that also contribute to increase the degree of the nodes. The exact contribution can be calculated by considering the different scenarios in which a 2-simplex $(i,j,k)$ can be attached to a node $i$ already having some links due to the first phase of the RSC construction. More precisely, the degree $k_i$ of node $i$ is incremented by 2 for each 2-simplex $(i,j,k)$ such that neither the link
$(i,j)$ nor the link $(i,k)$ are already present; this happens with probability $(1-p_1)^2$. Analogously, if either the link $(i,j)$ is already present 
but not $(i,k)$, or vice-versa, the addition of the 2-simplex $(i,j,k)$ increases the degree of $i$ by $1$. Since each case happens with the same probability $p_1(1-p_1)$ the contribution is therefore $2p_1(1-p_1)$. Overall, 
the degree $k_i$ increases on average 
by $2(1-p_1)$ for each
$2$-simplex attached to $i$.
Finally, for $p_1, p_\Delta \ll 1$, we can thus write the expected average degree $\langle k \rangle$ as the sum of the 
two contributions coming from the links and
the $2$-simplices, namely $\langle k \rangle \approx (N-1)p_1 + 2\langle k_{\Delta} \rangle (1-p_1)$. For any given size $N$, we can thus produce simplicial complexes having desired values of $\langle k \rangle$  and $\langle k_{\Delta} \rangle$ by fixing $p_1$ and $p_{\Delta}$ as:
\begin{subequations}\label{eq:ps}
	\begin{align}
	p_1 &= \frac{\langle k \rangle - 2\langle k_{\Delta} \rangle}{(N-1) -2\langle k_{\Delta} \rangle} \\
	p_{\Delta} &= \frac{2\langle k_{\Delta} \rangle}{(N-1)(N-2)}   \   .
	\end{align}
\end{subequations}

\subsection{Data availability.}
The SocioPatterns data sets were downloaded from \hyperref[https://www.sociopatterns.org/datasets]{sociopatterns.org/datasets}. 

\subsection{Code availability.}
The code and datasets are available at: \hyperref[https://github.com/iaciac/simplagion]{github.com/iaciac/simplagion}

\end{methods}

\begin{addendum}

 \item I. I. and V. L. acknowledge support from EPSRC Grant EP/N013492/1. I. I. acknowledges support from The Alan Turing Institute under the EPSRC Grant No. EP/N510129/1. G. P. acknowledges support from ADnD Grant by Compagnia San Paolo and from Intesa Sanpaolo Innovation Center. 
 
 \item[Author contributions] I.I., G.P., A.B. and
 V.L. designed the study. I.I. and G.P. performed the numerical analysis. I.I., G.P., A.B. and V.L. wrote the paper.
 
 \item[Competing interests] The authors declare that they have no competing interests. The funders had no role in study design, data collection and analysis, decision to publish, or preparation of the manuscript.

 \item[Correspondence] Correspondence and requests for materials
should be addressed to V.L.~(email: v.latora@qmul.ac.uk).
\end{addendum}

\clearpage

\begin{center}
	{\Large{\textbf{Supplementary Information \\ \bigskip
				``Simplicial models of social contagion"}}}\\ \bigskip
	\large{ Iacopini et al.}
	
	\setcounter{section}{0}
	\setcounter{table}{0}
	\setcounter{equation}{0}
	\setcounter{figure}{0}
	
	\makeatletter
	\renewcommand{\thefigure}{S\arabic{figure}}

\end{center}

\clearpage
\newpage

\section{Generalized degree distributions of empirical and synthetic simplicial complexes}

\begin{figure}[thb]
	\centering
	\includegraphics[width=0.9\textwidth]{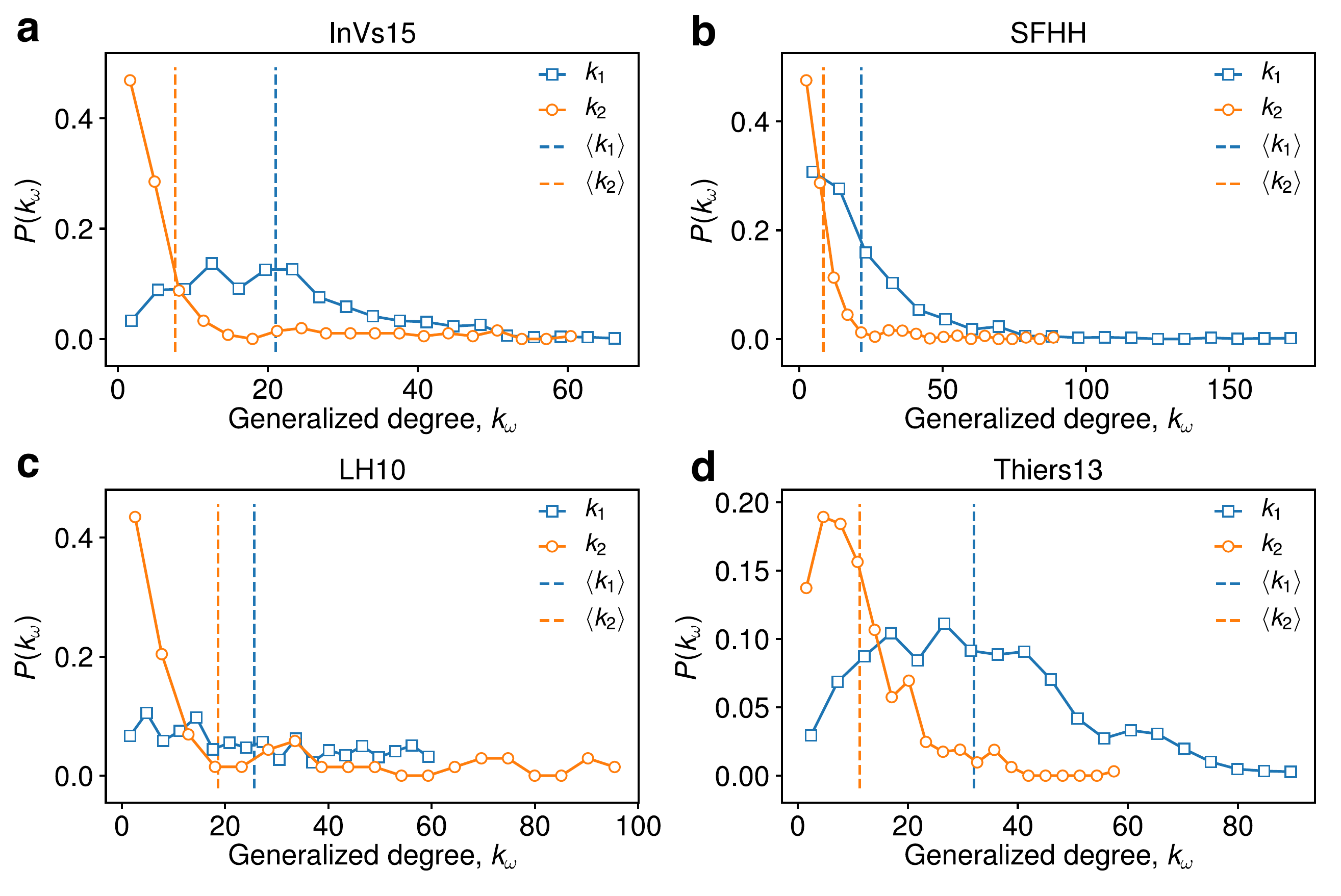}
	\caption{\label{fig:k_dist} Generalised degree distributions of random simplicial complexes created from real world data sets (see the data processing method described in the ``Methods" section of the main text). The four panels correspond to different social contexts, namely ({\bf a}) a workplace (InVS15), ({\bf b}) a conference (SFHH), ({\bf c}) a hospital (LH10) and ({\bf d}) a high school (Thiers13). The generalised degrees $k_1$ and $k_2=k_{\Delta}$ denote respectively the number of 1-simplices (blue) and 2-simplices (orange) incident in a node. The vertical dashed lines
		indicate the corresponding average values.}
\end{figure}

\begin{figure}[thb]
	\centering
	\includegraphics[width=0.4\textwidth]{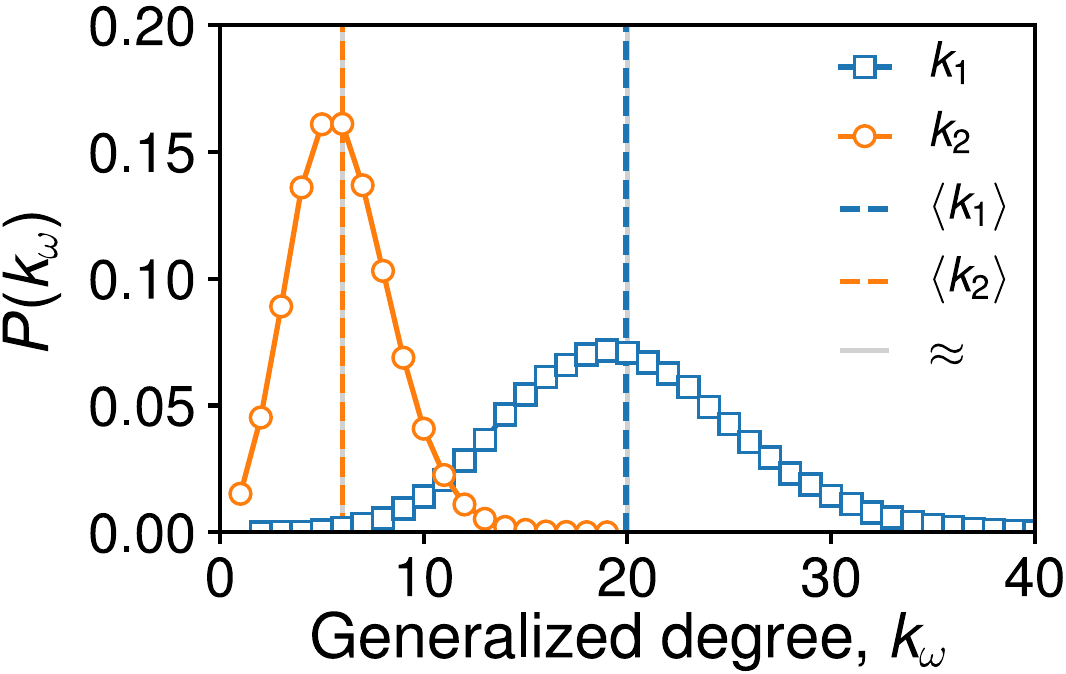}
	\caption{\label{fig:k_dist} Generalised degree distributions of random simplicial complexes (RSC) generated by the model described in the main text. The generalised degrees $k_1$ and $k_2=k_{\Delta}$ denote respectively the number of 1-simplices (blue) and 2-simplices (orange) incident in a node. The vertical lines compare the average values of $\langle k_1 \rangle$ and $\langle k_2\rangle$ obtained from multiple realizations of the model (coloured dashed lines) with the approximated values (continuous grey lines) calculated as described in the main text. }
\end{figure}

\clearpage
\newpage

\section{Hysteresis and system size}

\begin{figure}[thb]
	\centering
	\includegraphics[width=0.75\textwidth]{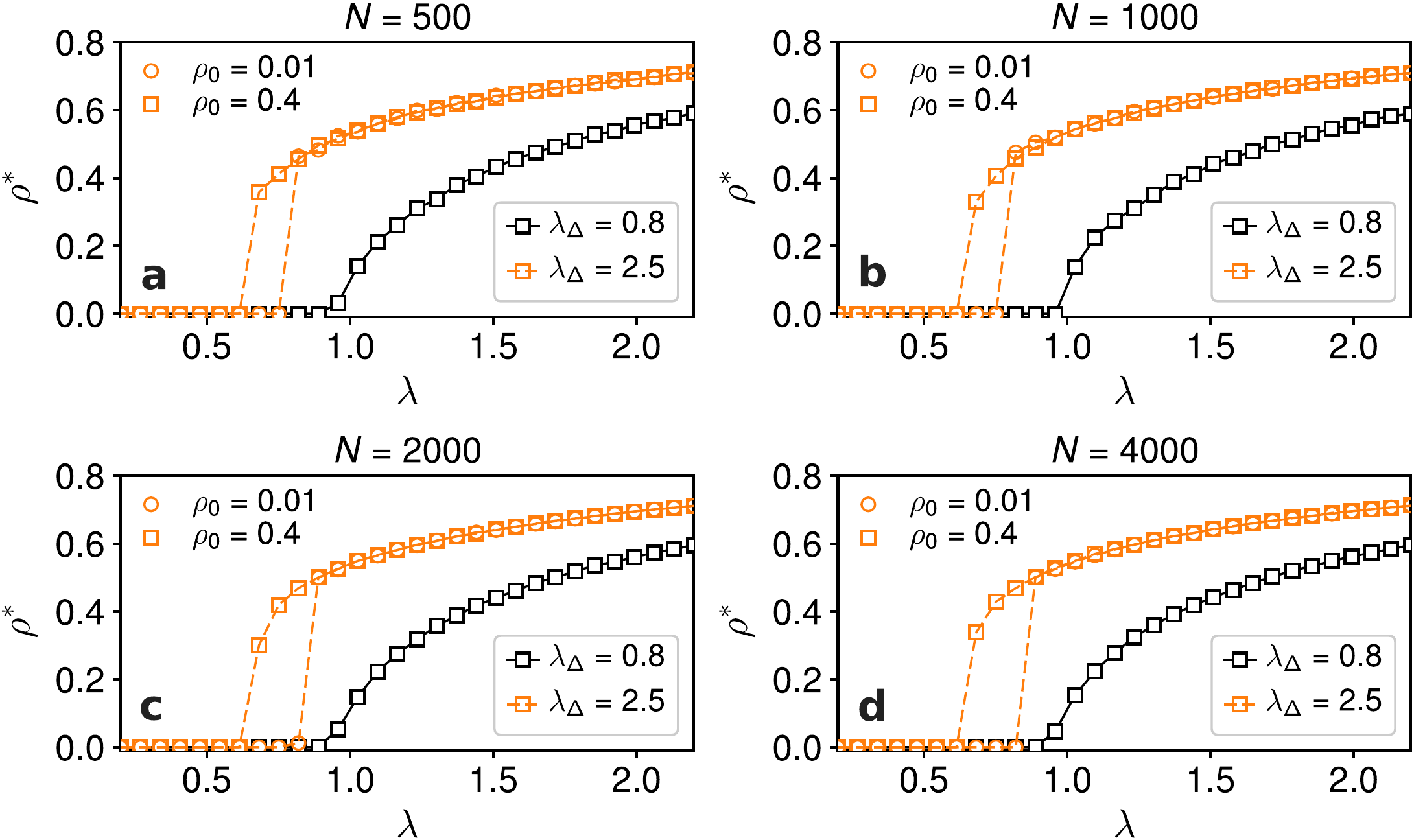}
	\vspace{-2em}
	\caption{\label{fig:hysteresis_size} Numerical exploration of the finite size effects
		on the hysteresis for a SCM of order $D=2$ on synthetic random simplicial complexes (RSC). The RSCs are generated with the procedure described in the main text, with parameters $p_1$ and $p_{\Delta}$ tuned in order to produce simplicial complexes with $\langle k \rangle \sim 20$ and $\langle k_{\Delta} \rangle \sim 6$. Different panels correspond to different system sizes, namely ({\bf a}) $N=500$, ({\bf b}) $N=1000$, ({\bf c}) $N=2000$, and ({\bf d}) $N=4000$. Each panel shows the average stationary fraction of infected individuals plotted against the rescaled infectivity $\lambda=\beta\langle k \rangle/\mu$. The parameter $\lambda_{\Delta} = \beta_{\Delta}\langle k_{\Delta} \rangle/\mu$ is set to $\lambda_{\Delta}=2.5$, which corresponds to the case in which we observe a discontinuous transition, with the formation of a a bistable region where healthy and endemic states co-exist and a hysteresis appears. The two types of orange symbols correspond to two different values of the initial density of infected individuals for $\lambda_{\Delta}=2.5$, namely $\rho_0=0.01$ (circles) and $\rho_0=0.4$ (squares). 
		The case $\lambda_{\Delta}=0.8$, in which we observe a continuous transition with no hysteresis, is shown for reference (black squares).}
	\vspace{-6em}
\end{figure}

\clearpage
\newpage

\begin{figure}[thb]
	\centering
	\includegraphics[width=0.75\textwidth]{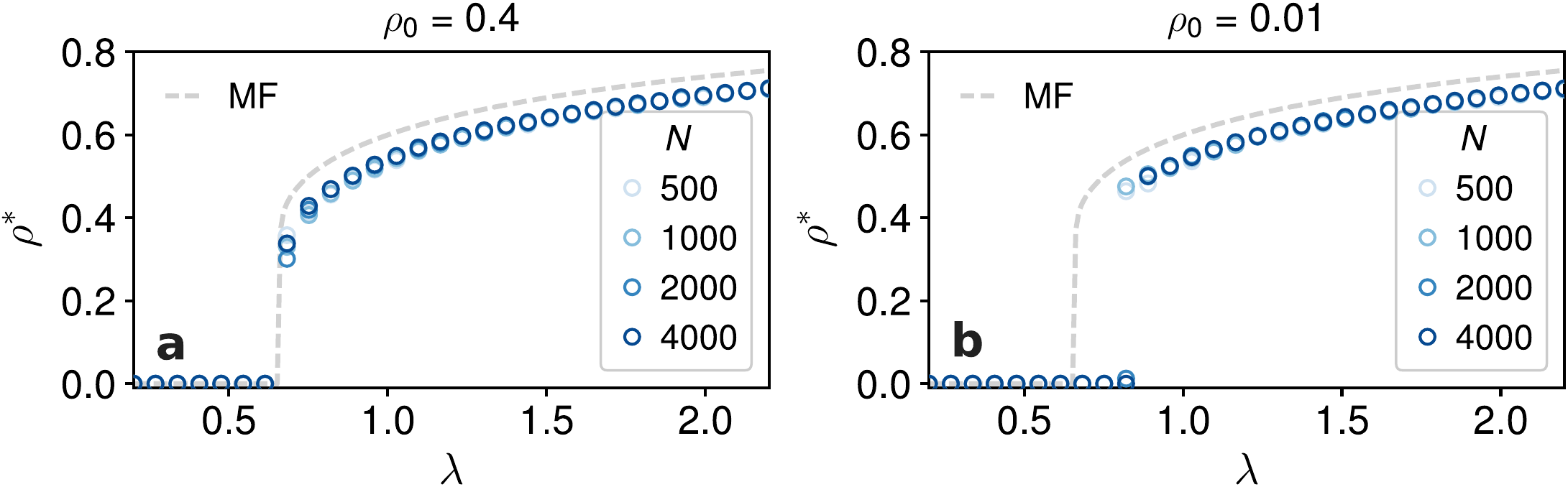}
	\caption{\label{fig:hysteresis_size_2} Numerical exploration of the finite size effects on the hysteresis for a SCM of order $D=2$ on synthetic random simplicial complexes (RSC). The two panels refer to two  
		different values of the initial density of infected individuals, namely ({\bf a}) $\rho_0=0.4$ and ({\bf b}) $\rho_0=0.01$. The dashed
		line corresponds to the mean-field result.}
\end{figure}

\clearpage
\newpage

\section{Cases of higher dimensions}

\section*{Case $D=3$}

Let us consider here a system with maximum dimension of simplices $D=3$. In this case the model has three spreading parameters $\beta_1$, 
$\beta_2 = \beta_\Delta$ and $\beta_3$, and the evolution equation  for $\rho(t)$ reads
\begin{equation}
d_t \rho(t) = - \mu \rho(t) + \beta \langle k \rangle \rho(t)(1 - \rho(t)) +  \beta_2 \langle k_2 \rangle \rho(t)^2 (1 - \rho(t)) +
\beta_3 \langle k_3 \rangle \rho(t)^3 (1-\rho(t))  .
\end{equation}
Finding the roots of $d_t \rho(t) = 0$ yields a polynomial of degree $3$, so it is possible to write these roots, corresponding to stable and unstable
fixed points of the dynamics, as functions of the parameters of the model. The process is however lengthy and cumbersome, and depends moreover on three parameters, so that
the representation of the whole phase diagram is not convenient.

As we want here simply to show that the phenomenology of the appearance of first order transitions obtained in the case $D=2$, is also observed in higher 
dimensions, we restrict ourselves 
for simplicity to the case $ \beta_\Delta = 0$, in which we will see that we can avoid writing the explicit solutions and resort instead to a graphical
solution. This case corresponds to the hypothesis
that contagion can occur only either through simple contagion or through cliques of size $4$ in which $3$ of the nodes are already infectious, and the 
evolution equation reduces to: 
\begin{equation}
d_t \rho(t) = - \mu \rho(t) + \beta \langle k \rangle \rho(t)(1 - \rho(t)) + \beta_3 \langle k_3 \rangle \rho(t)^3 (1-\rho(t))  .
\end{equation}
Setting 
$\lambda = \beta \langle k \rangle/\mu$,
$\lambda_3 = \beta_3 \langle k_3 \rangle / \mu$ and rescaling time by $\mu$ we obtain:
\begin{equation}
d_t \rho(t)  = \rho(t) (1-\rho(t)) \left( \lambda + \lambda_3 \rho^2 - \frac{1}{1 - \rho(t)} \right) 
\end{equation}
where we can define the functions $f_1(\rho)= \lambda + \lambda_3 \rho^2$ and $f_2(\rho)= 1/(1-\rho)$. The sign of the temporal evolution of the density of infectious is
thus given by the sign of the difference between $f_1 - f_2$. Note that $\rho(t)$ is by definition between $0$ and $1$ so we need to consider $f_1$ and
$f_2$ only between these limits. In this interval, $f_1$ is positive and increases monotonically from $\lambda$ for $\rho=0$ to $\lambda + \lambda_3$ for $\rho=1$. Function 
$f_2$ is also positive and strictly increasing, with $f_2(0)=1$ and $f_2$ diverging towards $+\infty$ as $\rho \to 1^-$.
We also note that the equation $f_1(\rho) = f_2(\rho)$ yields a polynomial of degree $3$, so it has at most $3$ real roots. 

Let us first consider the case $\lambda > 1$. Then at $\rho = 0$ we have $f_1 > f_2$, and as $\rho \to 1$, $f_1$ becomes smaller than $f_2$. Therefore, at small $\rho$, $d_t \rho$ is  positive and hence the state $\rho=0$ is unstable. 
More in detail, there are two possibilities:
\begin{itemize}
	
	\item  either there is one single crossing
	point of $f_1$ and $f_2$, at $\rho^*$. Then,  $d_t \rho(t) > 0$ if $\rho(t) < \rho^*$ and $d_t \rho(t) < 0$ if $\rho(t) > \rho^*$: for any $\rho(t=0) > 0$, the system goes to the stationary state
	$\rho(t \to \infty) = \rho^*$. This is similar to the usual SIS case with $\lambda_3 = 0$: the effect of a non-zero
	value of $\lambda_3$ is simply to shift the value of $\rho^*$.
	
	\item or
	there are three crossing points $\rho_1 < \rho_2 < \rho_3$.
	This occurs for certain combinations of values of $\lambda$ and $\lambda_3$. Then for $\rho(t) < \rho_1$, $d_t \rho(t) > 0$ so the absorbing state $\rho=0$ is again unstable.
	The state $\rho_2$ is also seen to be unstable while there are two stable fixed points $\rho_1$ and $\rho_3$: depending on the value of $\rho(t=0)$, the system
	will converge to one of these values. 
\end{itemize}
Hence, for $\lambda > 1$, the system always reaches a stationary state with a finite fraction of infectious nodes, which in some regions of the ($\lambda$,$\lambda_3$)
phase diagram, can depend on $\rho(t=0)$.

\medskip

\begin{figure}[thb]
	\centering
	\includegraphics[width=0.6\textwidth]{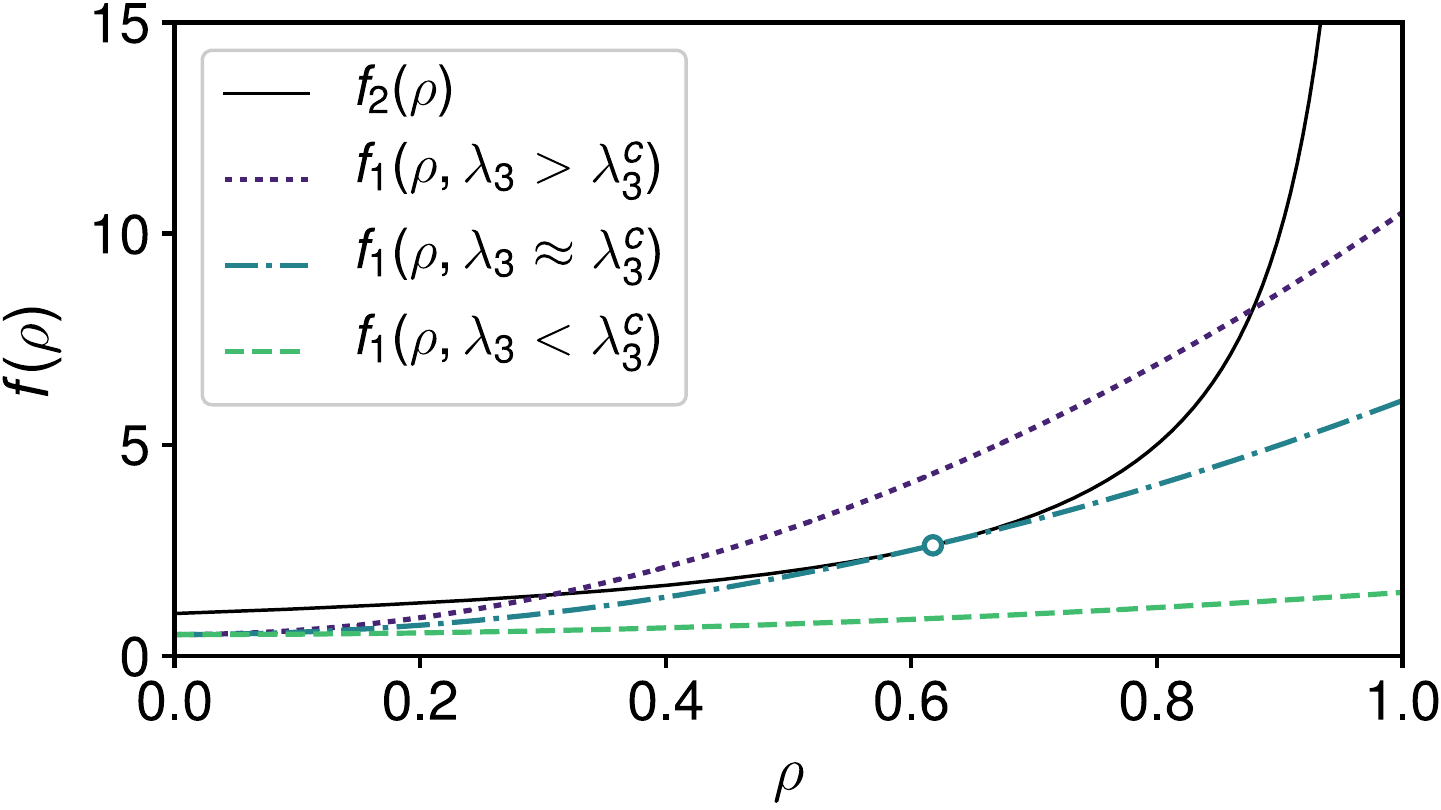}
	\caption{SCM of order $D=3$, case $\lambda = 0.5$, $\lambda_2=0$: $f_1(\rho)$ for various $\lambda_3$ ($<$, $\approx$ and $>\lambda_3^c$), and $f_2(\rho)$.
		$f_1$ is below $f_2$ both at $\rho=0$ and as
		$\rho \to 1$. The two curves therefore either do
		not cross (for $\lambda_3 < \lambda_3^c$),
		are tangent in $\rho_+ = \rho_-$ (for $\lambda_3 = \lambda_3^c$) or cross in two points 
		$\rho_-$ and $\rho_+$ 
		(for $\lambda_3 > \lambda_3^c$).}
	\label{fig:f1f2}
\end{figure}

Let us now consider the more interesting case $\lambda < 1$. Then $f_1(\rho) < f_2(\rho)$ both for $\rho=0$ and as $\rho \to 1$. 
Hence $f_1 - f_2$ is negative both in $0$ and $1$, and either $0$ or $2$ of the
roots of the equation $f_1(\rho) = f_2(\rho)$ are between $0$ and $1$.
Hence, for $\rho \in [0,1]$, either $f_1$ is always below $f_2$, or the two functions intersect in $2$ points that we call $\rho_-$ and $\rho_+$ ($\rho_- < \rho_+$): 
\begin{itemize}
	\item in the former case ($f_1(\rho) < f_2(\rho)$ $\forall \rho \in[0,1]$), $d_t \rho(t)$ is always negative so the only stationary state is the absorbing one $\rho = 0$;
	
	\item in the latter case, $d_t \rho$ is positive for $\rho(t)$ between $\rho_-$ and $\rho_+$ and negative else, so that
	\begin{itemize}
		\item if $\rho(t=0) < \rho_-$, $d_t \rho$ is negative, hence $\rho(t)$ decreases and the system converges to $\rho=0$
		
		\item if $\rho(t=0) > \rho_-$, the system converges towards $\rho(t \to \infty) = \rho_+ > 0$.
	\end{itemize}
	
\end{itemize}
At fixed $\lambda < 1$, the former case is obtained at small values of $\lambda_3$, while the latter is obtained for $\lambda_3$ large enough.
The situation is illustrated in Fig. \ref{fig:f1f2} for $\lambda = 0.5$.
At the transition $\lambda_3 = \lambda^c_3$ between these two cases, $\rho_- = \rho_+ > 0$ (the functions $f_1$ and $f_2$ 
are tangent in this point): the transition from 
$\rho(t \to \infty) = 0$ for $\lambda_3 < \lambda_3^c$ to 
$\rho(t \to \infty) = \rho_+$ (if $\rho(t=0) > \rho_-$) for $\lambda_3 > \lambda_3^c$ 
is thus a discontinuous one, in a similar way
to the case $D=2$ discussed in the main text.

\clearpage
\newpage

\section*{General $D$, with $\beta_1 = \cdots = \beta_{D-1}=0$}

For general $D$, there is no analytical solution for the stationary values of the density of infectious nodes. We show here however that, if
we consider that contagion can occur only through cliques of size $D+1$, i.e., if all spreading rates $\beta_1$, $\beta_2$, \dots, $\beta_{D-1}$
are null, there exists a discontinuous transition between the phase in which the spreading vanishes at low $\beta_D$ and the phase in which 
$\rho(t \to \infty)$ is finite at large $\beta_D$.

The evolution equation for $\rho$ reads
\begin{equation}
d_t \rho(t) = - \mu \rho(t) + \beta_D \langle k_D \rangle \rho(t)^D (1-\rho(t))  .
\end{equation}
Defining $\lambda_D = \beta_D \langle k_D \rangle / \mu$ and rescaling time by $\mu$ we obtain
\begin{equation}
d_t \rho(t)  = - \ \rho(t) \Big[ 1 - \lambda_D \rho^{D-1}(t) ( 1 - \rho(t)) \Big] .
\end{equation}
Defining $F_D (\rho) = 1 - \lambda_D \rho^{D-1} ( 1 - \rho )$, we see that the sign of $d_t \rho(t)$ is opposite to the sign of $F_D (\rho(t))$, so that
we need to study the sign of the function $F_D(\rho)$ for $\rho \in [0,1]$ (as the density $\rho(t)$ is by definition between $0$ and $1$).

We have $F_D(0)=F_D(1)=1$. Moreover, the derivative of $F_D$ is 
$$
F_D'(\rho) = \lambda_D (D \rho^{D-1} - (D-1) \rho^{D-2}) = 
D \lambda_D \rho^{D-2} (\rho - (1-1/D)).
$$
It is thus negative for $\rho < 1 -1/D$ and positive for $\rho > 1-1/D$: $F_D$ first decreases as $\rho$ increases, reaches a minimum at $\rho=1-1/D$
and then increases back to $1$ as $\rho$ increases to $1$. We have thus two cases:
\begin{itemize}
	\item if the minimum, $F_D(1-1/D)$, is positive, then $F_D(\rho) > 0$ for $\rho \in [0,1]$: therefore, $d_t \rho(t)$ is always negative for any $\rho(t) > 0$: the density
	of infectious nodes can only decrease and the contagion-free state $\rho=0$ is the only stable state. 
	
	\item if instead $F_D(1-1/D) < 0$, then, as $F_D(0)=F_D(1)=1$, by continuity the equation $F_D(\rho)=0$ has two roots in $[0,1]$, which we call $\rho_-$ 
	and $\rho_+$ ($\rho_- < \rho_+$). $F_D(\rho)$ is positive for $\rho \in [0,\rho_-)$ and $\rho \in (\rho_+,1]$ and negative between the two roots.
	Therefore 
	\begin{itemize}
		\item if $\rho(t=0) < \rho_-$, $d_t \rho(t=0)$ is negative, hence $\rho(t)$ decreases and the system converges to $\rho=0$
		
		\item if $\rho(t=0) > \rho_-$, the system converges towards $\rho(t \to \infty) = \rho_+ > 0$.
	\end{itemize}
	
	The condition to have $F_D(1-1/D) < 0$ and hence a non-trivial stationary state can be written simply as
	$$
	1- \lambda_D (1-1/D)^{D-1} (1/D) < 0
	$$
	i.e.,
	$$
	\lambda_D > \lambda_D^c = \frac{D^D}{(D-1)^{D-1}} .
	$$
	Note that for $\lambda_D = \lambda_D^c$, $\rho_-=\rho_+ = 1-1/D$ is strictly positive, showing that the transition at $\lambda_D^c$ is discontinuous.
	
	This shows therefore that for $\beta_1 = \cdots = \beta_{D-1}=0$, we have the same phenomenology for any $D$ as for the case $D=2$ studied
	on the main text: a discontinuous transition occurs at $\lambda_D^c = \frac{D^D}{(D-1)^{D-1}}$ between an absorbing state $\rho=0$
	and a stationary state with a non-zero density of infectious individuals $\rho_+ > 0$.

\end{itemize}

\clearpage
\newpage

\section{Hypergraphs and simplicial complexes}

Hypergraphs are a generalization of the concept of  graphs in which the edges, called  
hyperedges, can join any number of vertices. Formally, a hypergraph $\cal H$ is the pair of sets $(V ,E)$, where 
$V$ is a set of vertices, and the set of hyperedges $E$ 
is a subset of the power set  $P(V)$ of $V$. 
Simplicial complexes are therefore special kinds of hypergraphs, which contain all subsets of every hyperedge. A simplicial complex $\cal K$ on the set of vertices $V$ can indeed be seen as a hypergraph 
$\cal H$ on $V$ if the latter satisfies the extra requirement that, for each $\sigma \in E$, and for all $ \nu \neq \emptyset$ such that $\nu\subseteq\sigma$, 
we also have $\nu\in E$. 
Such an extra requirement seems 
appropriate in the context of models of social 
interactions considered in our work,  
and it also turns useful to keep the model simple 
and amenable to analytical solution. 
However, the SCM can be straightforwardly extended to model the more general case of complex contagion processes on hypergraphs.

\clearpage
\newpage

\section{Results on empirical simplicial complexes without data augmentation}

\begin{figure}[thb]
	\centering
	\includegraphics[width=\textwidth]{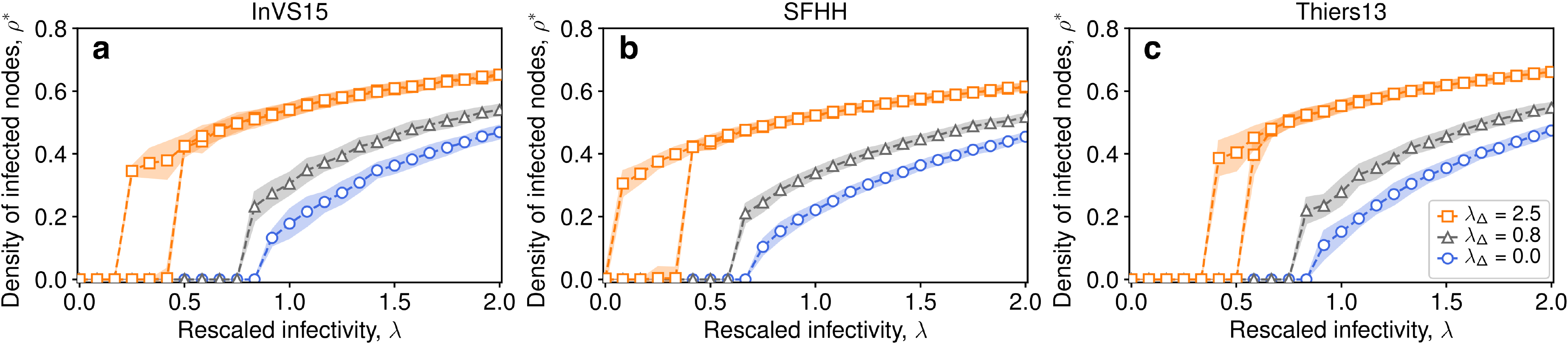}
	\caption{\label{fig:SI_sociopatterns} SCM of order $D=2$ on real-world higher-order social structures without data augmentation. Simplicial complexes are constructed from high-resolution face-to-face contact data recorded in a workplace ({\bf a}), a conference ({\bf b}), and a high school ({\bf c}). The average fraction of infected nodes in the stationary state obtained 
		numerically is plotted against the rescaled infectivity $\lambda=\beta\langle k \rangle/\mu$ for $\lambda_{\Delta}=0.8$ (black triangles) and $\lambda_{\Delta}=2.5$ (orange squares). The blue circles denote the simulated curve for the  standard SIS model ($\lambda_{\Delta}=0$), which does not consider higher order effects. For $\lambda_{\Delta}=2.5$ a bi-stable region appears, where healthy and endemic states co-exist.}
\end{figure}

\end{document}